\documentclass[12pt]{article}
\usepackage{epsfig,amssymb,cite,multirow} 
\usepackage{amsmath}

\usepackage{amscd}
\usepackage{amsmath,graphicx}
\usepackage{xypic}
\usepackage[matrix,arrow,curve]{xy}
\usepackage{graphicx}
\usepackage{verbatim}
\usepackage{epsf}
\usepackage{latexsym}
\usepackage{amsmath,amsfonts,amssymb,amsthm}
\usepackage{amsmath,amsthm}

\def\bseq{\begin{subequation}}  
\def\eseq{\end{subequation}}
\def\bsea{\begin{subeqnarray}}  
\def\esea{\end{subeqnarray}}

\newcommand{\bbox}{\lower.2ex\hbox{$\Box$}}

\newcommand{\beq}{\begin{equation}}
\newcommand{\eeq}{\end{equation}}
\newcommand{\bea}{\begin{eqnarray}}
\newcommand{\eea}{\end{eqnarray}}
\newcommand{\ena}{\end{eqnarray}}

\newcommand{\Tr}{{\rm Tr}}
\renewcommand{\(}{\left(}
\renewcommand{\)}{\right)}
\renewcommand{\[}{\left[}
\renewcommand{\]}{\right]}

\newcommand{\be}{\begin{equation}}
\newcommand{\ee}{\end{equation}}

\begin{document}
\setcounter{page}{0}
\begin{titlepage}
\titlepage
\begin{flushright}
\end{flushright}
\begin{center}
\LARGE{\Huge
Non-supersymmetric CS-matter theories with known AdS duals
\\}
\LARGE{\Huge  }
\end{center}
\vskip 1.5cm \centerline{{\bf Davide Forcella$^{a}$ \footnote{\tt dforcell@ulb.ac.be}
  and Alberto Zaffaroni$^{b}$ \footnote{\tt alberto.zaffaroni@mib.infn.it}
}}
\vskip 1cm
\footnotesize{

\begin{center}

$^a$ Physique Th\'eorique et Math\'ematique and International Solvay Institutes\\
Universit\'e Libre de Bruxelles, C.P. 231, 1050Bruxelles, Belgium \\
\medskip

$^b$ Dipartimento di Fisica, Universit\`a di Milano--Bicocca, I-20126 Milano, Italy\\
            and\\
            INFN, sezione di Milano--Bicocca,
            I-20126 Milano, Italy\end{center}}

\bigskip

\begin{abstract}

We consider three dimensional conformal field theories living on a stack of N anti-M2 branes at the tip of eight-dimensional supersymmetric cones. The corresponding supergravity solution is obtained by changing sign to the four-form in the Freund-Rubin solution representing  M2 branes  (``skew-whiffing''  transformation)  and it is known to be stable.  The existence of these non supersymmetric, stable field theories, at least in the large N limit, is a peculiarity of the AdS$_4$/CFT$_3$ correspondence with respect to the usual AdS$_5$/CFT$_4$, and it is worthwhile to study it. We analyze in detail  the KK spectrum of the skew-whiffed solution associated with $S^7/\mathbb{Z}_k$ and we speculate on the natural field content for a candidate non-supersymmetric dual field theory.

\vskip 1.5truecm
\noindent Contribution prepared for the AHEP special issue on "Computational Algebraic Geometry in String and Gauge Theory" 

\end{abstract}

\vfill
\begin{flushleft}
{\today}\\
\end{flushleft}
\end{titlepage}

\newpage


\section{Introduction}

There has been some progress in understanding the conformal field theories living on a stack of N M2 branes at the tip of non compact eight-dimensional cones. These theories are  holographically dual to  Freund-Rubin compactifications of M theory of the form AdS$_4\times$ H$_7$, where H$_7$ is the  Einstein manifold at the base of the cone.  In the case of large
supersymmetry, ${\cal N}\ge 3$, the conformal field theory has been identified with a Chern-Simons theory in \cite{BGL, Aharony:2008ug, Benna:2008zy, Hosomichi:2008jb, Jafferis:2008qz, Aharony:2008gk}. The  case where the cone is a Calabi-Yau four-fold corresponds to ${\cal N}=2$ 
supersymmetry, and a general construction of the dual Chern-Simons theories 
has been discussed in \cite{Martelli:2008si,Hanany:2008cd};
a large number of models have been subsequently constructed  \cite{Hanany:2008fj,Ueda:2008hx, Imamura:2008qs, Franco:2008um, Hanany:2008gx, Davey:2009sr, Franco:2009sp,Amariti:2009rb, Martelli:2009ga, Benini:2009qs, Jafferis:2009th}. For the smallest amount of supersymmetry, ${\cal N}=1$, the cone is a Spin(7) manifold and a generic construction of the dual field theories for orbifolds has been given in \cite{Forcella:2009jj}.
In this paper we are interested in the three dimensional conformal field theories living on a stack of N anti-M2 branes at the tip of eight-dimensional cones with at least Spin(7) holonomy. The existence of these non supersymmetric, stable field theories, at least in the large N limit, is a peculiarity of the AdS$_4$/CFT$_3$ correspondence with respect to the usual AdS$_5$/CFT$_4$, and it is worthwhile to study it\footnote{Previous studies of three dimensional non-supersymmetric field theories in similar setups can be found in \cite{Armoni:2008kr, Narayan:2009uy}.}.
   
The supergravity solutions of N anti-M2 branes at the tip of supersymmetric cones is obtained by changing sign to the four-form in the familiar solutions representing  M2 branes. This operation is called  ``skew-whiffing'' in the supergravity literature and it is proven to produce a stable non-supersymmetric background \cite{Duff:1984sv}. The KK spectrum of the final theory is obtained from the original spectrum by a reshuffling of states.
We will mostly focus on the simple case H$_7=S^7$ in this paper. In this particular case  the skew-whiffing procedure is equivalent to a triality transformation in the isometry group $SO(8)$ and it leaves the theory invariant. However, following the recent  results on AdS$_4\times $CFT$_3$, we are really interested in H$_7=S^7/\mathbb{Z}_k$. For $k\ge 2$ the skew-whiffing produces a stable non-supersymmetric background. It is interesting then
to find a plausible dual field theory\footnote{In \cite{Imaanpur:2010yk} there is a discussion of the effect of the skew-whiffing on the dual field theory for $k=1$.}. In this paper we analyze in details the KK spectrum of the skew-whiffed theory
and we speculate on a candidate for the underlying  field theory. Without supersymmetry, it is difficult to make
explicit checks of the proposal. We are guided in our search by the attempt of implementing the skew-whiffing
procedure to the ABJM theory \cite{Aharony:2008ug}.

The paper is organized as follows. In Section \ref{SS} we discuss the skew-whiffing construction. In Section \ref{SSABJM} we discuss a natural field content  for the skew-whiffing of the ABJM theory.
In Section \ref{checks} we make some checks of the proposal and in particular we see that the relevant features of the KK spectrum in supergravity are compatible with the operator content of the  field theory. In Section \ref{conc} we discuss possible generalizations.

\section{The skew-whiffed solutions}\label{SS}

 M-theory has a very natural compactification to four dimensions \footnote{For a comprehensive review see \cite{Duff:1986hr}.}. It comes from
 the Freud-Rubin ansatz for the 3-form field,
\begin{equation}
F_{\mu\nu\rho\sigma}= 3 m \epsilon_{\mu\nu\rho\sigma}\, ,  \qquad  \qquad F_{mnpq}=0
\end{equation}
where $m$ is a real constant, the greek letters label the four space-time directions, while the latin letters label the internal seven dimensional 
space. As a consequence the metric satisfies the equations,
\begin{equation}
R_{\mu\nu}= - 12 m^2 g_{\mu\nu}\, ,  \qquad \qquad R_{mn}= 6 m^2 g_{mn}
\end{equation}
and M theory spontaneously compactifies to AdS$_4$ $\times$ H$_7$, where H$_7$ is an Einstein space. If the Einstein 
manifold H$_7$ has at least weak G$_2$ holonomy, then the solution is supersymmetric and hence stable. This solution 
can indeed be interpreted as the near horizon solution for M2 branes at the tip of the real cone C(H$_7$).
There is another very natural solution that can be obtained sending $m$ to $-m$: 
the four form changes sign, while the metric is invariant. 
This solution can be interpreted as the near
horizon solution for anti-M2 branes at the tip of the real cone C(H$_7$) and it is called the ``skew-whiffed'' solution.
 
It is interesting to observe that the skew-whiffed transformation changes 
the sign of the Page charge $P=1/\pi^4 \int_{H_7} * F$, and can indeed be 
equivalently interpreted as a change of orientation for H$_7$.

The number of preserved supersymmetries of a solution is the number of 
independent Killing spinors $\eta$ solving the equation
\begin{equation}
(\nabla_n -\frac{1}{2} m \Gamma_n)\eta = 0
\end{equation}
This supersymmetric condition explicitly depends on $m$. Indeed it is possible to prove that,
 with exception of the round sphere $S^7$, where both orientations give the maximum 
supersymmetry $\mathcal{N}=8$, at most one orientation can have $\mathcal{N}>0$  \cite{Duff:1986hr}. 
Moreover it is possible to prove that the skew-whiffing of  a supersymmetric 
Freud-Rubin solution is perturbatively stable \cite{Duff:1984sv}.
Hence, given a supersymmetric Freud-Rubin solution, we can find  another 
non supersymmetric stable solution, obtained by applying the skew-whiffing transformation\footnote{See \cite{Denef:2009tp, Gauntlett:2009bh} for some recent applications of the skew-whiffed solutions.}.

These theories are expected to have a well defined  dual non-supersymmetric three dimensional 
conformal field theory. In this paper we will discuss  a natural field content  for these theories. 

It is important to observe that the skew-whiffing procedure is somehow peculiar of
 M theory. Indeed the same transformation 
could be applied to the five form in the Freud-Rubin ansatz for type IIB supergravity. 
However, in this case, the resulting theory would still be supersymmetric.

\section{The skew-whiffing of the ABJM theory}\label{SSABJM}


When the seven dimensional Einstein space is a $\mathbb{Z}_k$ quotient of the round sphere, 
S$^7$/$\mathbb{Z}_k$, the dual field theory is the well-known ABJM model \cite{Aharony:2008ug}.
This is an $\mathcal{N}=6$ three dimensional  U(N)$\times$U(N) Chern-Simons theory with  levels k and -k, coupled to four complex bosons $X^A$ and four complex fermions $\psi_A$ transforming in the bi-fundamental of the gauge group. The global symmetry is SU(4)$_R \times$U(1), the non-abelian  part being the R symmetry group, while the abelian part being the baryonic symmetry. The bi-fundamental bosons $X^A$ transform  in the representation ${\bf 4}_1$ of the global symmetry group and the  bi-fundamental fermions $\psi_A$ transform in the $\bar{{\bf 4}}_{-1}$. All fields have canonical dimension: $1/2$ for the bosons and $1$ for  the fermions.

This specific field content can be understood as  the decomposition of the degrees of freedom of the ${\cal N}=8$  conformal theory dual to $S^7$ under the breaking of the R symmetry group SO(8)$\rightarrow$SU(4)$\times$U(1) induced by  $\mathbb{Z}_k$. 
The natural field content of the theory  describing $S^7$ consists in  a scalar $\Phi_a$ transforming in the ${\bf 8}_v$ vectorial representation of SO(8)  and a fermion $Y_{\alpha}$ transforming in the ${\bf 8}_s$ spinorial representation of SO(8) \footnote{This is  the {\it singleton} representation of the superconformal group. };  the eight supercharges transform in the ${\bf 8}_c$ conjugate spinor 
representation. These are the degrees of freedom that we expect to live   on a supersymmetric M2 brane in flat space. The existing ${\cal N}=8$ Chern-Simons theory describing membranes, the BLG model \cite{BGL}, have indeed this field content. The ${\cal N}=8$ multiplet decomposes as: ${\bf 8}_v \rightarrow {\bf 4}_1 + {\bf \bar{4}}_{-1}, {\bf 8}_s \rightarrow {\bf 4}_1 + {\bf \bar{4}}_{-1}$ under SO(8)$\rightarrow$ SU(4)$\times$U(1) and we recover the ABJM matter content describing  $S^7/\mathbb{Z}_k$. The ABJM model has enhanced ${\cal N}=8$ supersymmetry for $k=1$ but  this enhancement is not manifest and it is conjectured to be realized through light monopole operators which become relevant for $k=1$.  

We would like to identify a candidate for the skew-whiffing solution on $S^7$/$\mathbb{Z}_k$. As we 
explained in the previous section, the skew-whiffing procedure can be understood as the change of
orientation of $S^7$, and hence as a parity transformation on 
$\mathbb{R}^8$. 
The scalar fields living on the M2 can be seen as the coordinates of the transverse space, and, under 
parity,  an odd number of them change sign, while the associated eight dimensional spinor changes chirality.

The skew-whiffing procedure can be seen  as a map,

\begin{equation}{\rm SW}: \qquad  ({\bf 8}_v\, , {\bf 8}_s) \qquad \rightarrow \qquad 
({\bf 8}_v\, , {\bf 8}_c)\, . \label{SWrep}\end{equation}

An explicit  realization of this map on an $\mathcal{N}=8$ multiplet is
\begin{equation}{\rm SW}: \qquad  (\Phi^i\, ,\Phi^8\, ,Y^{a}) \qquad \rightarrow \qquad 
(\Phi^i\, , - \Phi^8\, , (\Gamma^8)_{\dot{a} a} Y^{a})\, . \label{SW}\end{equation}

For $k=1$, this transformation should be an invariance of the dual field theory, since $S^7$ is skew-whiffing invariant. We can test this transformation in the BLG theory \cite{BGL}, which  is an explicit $\mathcal{N}=8$ lagrangian theory invariant under the SO(8) R symmetry and describes a pair of membranes. It is an easy exercise to see that the BLG Lagrangian is indeed invariant under the transformation (\ref{SW}), as required by the supergravity skew-whiffing transformation. In the ABJM theory with $k=1$, this transformation is, on the other hand, non locally realized.
 
For generic k,  the SO(8) symmetry is broken to SU(4)$\times$U(1). Under skew-whiffing, the ${\bf 8}_s$ fermions are replaced by ${\bf 8}_c$ and   the  fields decompose as: ${\bf 8}_v \rightarrow {\bf 4}_1 + {\bf \bar{4}}_{-1}$, ${\bf 8}_c \rightarrow {\bf 1}_2 + {\bf 6}_0 + {\bf 1}_{-2}$. The original  supersymmetry charges would now transform in the ${\bf 8}_s \rightarrow {\bf 4}_1 + {\bf \bar{4}}_{-1}$. None is invariant under the U(1) symmetry and hence the theory naturally breaks all the supersymmetries.

It is natural to expect that  the dual of the skew-whiffed S$^7$/$\mathbb{Z}_k$ is  a non-supersymmetric three dimensional Chern-Simons theory with matter. Let us try to understand what is the natural field content. We still expect a gauge group U(N)$\times$U(N) and a global symmetry SU(4).
Comparing with the geometric action of the SW transformation, we introduce   complex scalars $X^A$ transforming in the bi-fundamental of the gauge groups and in the fundamental of the global symmetry SU(4);  real fermions $\Psi^I$ transforming in the  adjoint of the first gauge group and  in the antisymmetric of SU(4); complex fermions  $\xi$ transforming in the bi-fundamental of the SU(N)$\times$SU(N) gauge group,  with charge 2 under the  difference of the two U(1) gauge factors and as singlets under the global symmetry SU(4).
The matter content of such a  theory is summarized  in  Table 1 below. 
\\
\begin{table}[h!]
\begin{center}
\begin{tabular}{|c|c|c|c|c|c|}
\hline
Fields & $\hbox{ SU(N)}_1$ & $\hbox{SU(N)}_2$ & $\hbox{U(1)}_1-\hbox{U(1)}_2$ & $\hbox{U(1)}_1+\hbox{U(1)}_2$ & $\hbox{SU(4)}$ \\
\hline
$X^A$ & ${\bf N}$ & ${\bf\bar{ N}}$ & ${\bf 1}$ & ${\bf 0}$ & ${\bf 4}$ \\
\hline
$\Psi^I$ & Adj & ${\bf 0}$ & ${\bf 0}$ & ${\bf 0}$ & ${\bf 6}$ \\
\hline
$\xi$ & ${\bf N}$ & ${\bf\bar{ N}}$ & ${\bf 2}$ & ${\bf 0}$ & ${\bf 1}$ \\
\hline
\end{tabular}
\label{quivSS}
\end{center}
\caption{Matter content and charges.}
\end{table}

As we will see in the next Section,  the supergravity KK spectrum  predicts the existence of operators with
integer dimensions in the dual field theory. A peculiarity of the skew-whiffing transformation indeed is the fact that it just reshuffles the KK states without changing their dimensions. To match the KK spectrum with the above field content  we need to assume that all the  fields have canonical dimensions: 1/2 for the scalars and 1 for the fermions.
It is then tempting  to write a classically conformal invariant Lagrangian for the fields $X^A,\Psi^I,\xi$.
The most general Lagrangian we can write is
\begin{equation}\label{SN3} 
\mathcal{L}= \mathcal{L}_{CS} + \mathcal{L}_{kin} + V_{bos}+ V_{fer}\ ,\nonumber
\end{equation}
where
\begin{eqnarray}\label{inter}
& & \mathcal{L}_{CS}=  \frac{k}{4\pi} \Tr \[ \epsilon^{\mu \nu \lambda} \( A_{ \mu} \partial_{\nu} A_{  \lambda} + \frac{2i}{3} A_{  \mu} A_{ \nu} A_{  \lambda} \) - \epsilon^{\mu \nu \lambda} \( \tilde{A}_{ \mu} \partial_{\nu} \tilde{A}_{  \lambda} + \frac{2i}{3} \tilde{A}_{  \mu} \tilde{A}_{ \nu} \tilde{A}_{  \lambda} \) \]\ ,  \nonumber \\
& & S_{kin}= \Tr \( - D^{\mu}X^A D_{\mu}X^{\dagger}_A + i \bar{\Psi}^I \gamma^{\mu}D_{\mu} \Psi^I + i \bar{\xi}^{\dagger} \gamma^{\mu}D_{\mu} \xi \)\ , \nonumber \\
& & V_{bos}= \frac{ \pi^2 }{12 k^2}\Tr \big( X^A X_A^{\dagger}X^B X_B^{\dagger}X^C X_C^{\dagger} +  X_A^{\dagger}X^A X_B^{\dagger}X^B X_C^{\dagger}X^C+ \nonumber\\
& & \hbox{   } \hbox{  } \hbox{   } \hbox{    }  \hbox{    } \hbox{    } \hbox{    } \hbox{    } \hbox{   } 4 X^A X_B^{\dagger}X^C X_A^{\dagger}X^B X_C^{\dagger} - 6 X^A X_B^{\dagger}X^B X_A^{\dagger}X^C X_C^{\dagger}\big)\ , \nonumber\\
& & V_{fer}= \frac{i \pi}{2 k}\Tr \( c_1 \(\Gamma^{IJ}\)^B_A X^A X^{\dagger}_B \Psi^I \Psi^J + c_2( 
X^A X^{\dagger}_A \xi \xi^{\dagger}- X^{\dagger}_A X^A \xi^{\dagger} \xi ) \)\ .
\end{eqnarray}
where the covariant derivatives act according to the Table 1. We keep the same bosonic potential  of the ABJM theory, and we introduce two real couplings $c_i$ to parametrize the fermionic potential.  The same potential of the
ABJM theory guarantees that a probe see $\mathbb{R}^8/\mathbb{Z}_k$ as a moduli space; in the skew-whiffed supergravity background an anti-M2 brane feels no potential, exactly as a M2 brane in the original background.
The Lagrangian is scale invariant at classical level. Obviously, without supersymmetry, there is no guarantee that the theory remains conformal invariant at the quantum level and, in general, we expect that the  fields acquire anomalous dimensions. It is tempting to  speculate that, in the limit where supergravity is valid (large $N$ limit and strong coupling $N/k\gg1$ in type IIA), the theory becomes conformal with canonical dimensions for the fields.

One could try to impose a relation between the $c_i$ in the fermionic potential by  requiring that the scalar BPS operators of the ABJM theory do not acquire anomalous dimensions at weak coupling in the SW ABJM theory. This constraint is suggested by the dimensions of the scalar part of the KK spectrum in supergravity, and the strong assumption that we can extrapolate from strong to weak coupling. The argument goes as follows. Starting from the lagrangian (\ref{inter}) it is possible to compute the quantum part of the two loop dilatation operator ( mixing matrix ) for small values of N/k. The mixing matrix acting on the gauge invariant operators of the field theory gives their anomalous dimensions.  The mixing matrix acting on the gauge invariant operators done by contracting only the scalar fields $X^A$ and $X^{\dagger}_A$ was computed in \cite{Minahan:2008hf} for the ABJM theory. 
The computation can be easily repeated for the SW ABJM theory: the only difference is the contribution coming from the fermions running into the loops, and in particular their contribution to  the identity and the trace part of the mixing matrix.  By imposing that the scalar mixing matrix of the SW ABJM theory is annihilate by the scalar symmetric traceless operators (\ref{scalarBPS}) we find  a constraint on the $c_i$. It would be much harder to perform a similar computation with operators with fermionic insertions to see if some of them remain with canonical  
 dimensions with such a choice of $c_i$.
 


\section{The KK spectrum}\label{checks}

In this section, we discuss in details the KK spectrum of the skew-whiffed M theory solution on $S^7/\mathbb{Z}_k$ and of its reduction to type IIA.  The KK states  should correspond to the single trace operators of the dual field theory with  finite dimensions in the large N and strong coupling limit. It obviously difficult to predict the spectrum of operators with finite dimensions in a non supersymmetric theory. However, the distinctive features of the KK spectrum  put constraints on the dual field content. 


 
 \subsection{The spectrum on $S^7$}
The KK spectrum of the M-theory Freud-Rubin solution on $S^7$ is well-known 
\cite{Sezgin:1983ik, Biran:1983iy} and reported for completeness in Table \ref{S7}.  The states are classified by  the dimension  and the representation under SO(8) and are organized in superconformal multiplets  specified by an integer number $m\ge 2$. The lowest state in each multiplet is scalar field of dimension $m/2$ transforming in the $[m,0,0,0]$ representation of SO(8). The other states are obtained by acting on the lowest state with the ${\cal N}=8$ supersymmetries transforming as ${\bf 8}_c=[0,0,0,1]$ under SO(8). The multiplet is short and has spin range equal to two, compared with the spin range of four
of a long multiplet. The multiplets $m=2,3$ sustain further shortenings; $m=2$ corresponds to the {\it massless} 
multiplet of the ${\cal N}=8$ gauged supergravity. Partition functions encoding the  spectrum of $S^7$ (and of its quotient) in a related context can be found 
in \cite{Hanany:2008qc,Hanany:2010zz,Bianchi:2010mg,Samsonyan:2011gw}.

\begin{table}[h!]\begin{center}
\begin{tabular}{|c| p{3cm} |c|}
\hline
Spin &  $\hbox{SO(8)}$ & $\Delta$  \\
\hline
$2^+$ & $[m-2,0,0,0]$ & $\frac{m+4}{2}$ \\
\hline
$\frac{3}{2}^{(1)}$ & $[m-2,0,0,1]$ & $\frac{m+3}{2}$ \\
\hline
$\frac{3}{2}^{(2)}$ & $[m-3,0,1,0]$ & $\frac{m+5}{2}$ \\
\hline
$1^{-(1)}$ & $[m-2,1,0,0]$ & $\frac{m+2}{2}$ \\
\hline
$1^{+(2)}$ & $[m-3,0,1,1]$ & $\frac{m+4}{2}$ \\
\hline
$1^{-(2)}$ & $[m-4,1,0,0]$ & $\frac{m+6}{2}$ \\
\hline
$\frac{1}{2}^{(1)}$ & $[m-1,0,1,0] $ & $\frac{m+1}{2}$ \\
\hline
$\frac{1}{2}^{(2)}$ & $[ m-3,1,1,0] $ & $\frac{m+3}{2}$ \\
\hline
$\frac{1}{2}^{(3)}$ & $[m-4,1,0,1] $ & $\frac{m+5}{2}$ \\
\hline
$\frac{1}{2}^{(4)}$ & $[m-4,0,0,1] $ & $\frac{m+7}{2}$ \\
\hline
$0^{+(1)}$ & $[m,0,0,0]$ & $\frac{m}{2}$ \\
\hline
$0^{-(1)}$ & $[m-2,0,2,0]$ & $\frac{m+2}{2}$ \\
\hline
$0^{+(2)}$ & $[m-4,2,0,0]$ & $\frac{m+4}{2}$ \\
\hline
$0^{-(2)}$ & $[m-4,0,0,2]$ & $\frac{m+6}{2}$ \\
\hline
$0^{+(3)}$ & $[m-4,0,0,0]$ & $\frac{m+8}{2}$ \\
\hline
\end{tabular}
\end{center}
\caption{The spectrum on $S^7$. } 
\label{S7}
\end{table}

We can understand the structure of the multiplets by considering a very simple theory of eight free bosons $\Phi^i$ and eight free fermions $Y^a$.
These are the expected degrees of freedom living on  a membrane in flat space. The lowest state of the $m$-th KK supermultiplet can be represented by 
$ \Phi^m$, which is a schematic expression for the product of $m$ fields completely symmetrized in their SO(8) indices and with all the traces removed:
\begin{equation}
  \Phi^m \equiv \Phi^{\{i_1}\cdots \Phi^{i_m\}} -{\rm traces}
\end{equation}
The other states are obtained by applying the  supersymmetry transformations that schematically read
\begin{eqnarray}
[ Q^{\dot a}_{ \alpha}, \Phi^i ] &=& i (\tilde\Gamma^{i})^{\dot a a} Y^{a}_{\alpha} \nonumber\\
\{Q^{\dot a}_{\alpha}, Y^{a}_{\beta} \}& =&  (\gamma^\mu)_{\alpha\beta} D_\mu \Phi^i (\Gamma^i)^{a\dot a} \end{eqnarray}
The first spinorial state  $\frac{1}{2}^{(1)}$  with dimension $\frac{m+1}{2}$ reads 
\begin{equation}
  Y^{a}_{\alpha}  \Phi^{m-1} \end{equation}
where, again,  $\Phi^m$ will represent a fully symmetrized and traceless string of fields $\Phi^{i_k}$. The indices are contracted in such a way that the operator transforms as $[m-1,0,1,0]$. 
The states with dimension  $\frac{m+2}{2}$ are obtained by applying two supercharges whose Lorentz indices can be antisymmetrized or symmetrized
giving a scalar $ 0^{-(1)}$ and a vector $1^{-(1)}$ of schematic form
\begin{equation}
 Y_{[ \alpha}^{\{a}Y_{\beta ]}^{b\}}   \Phi^{m-2} \, , \,\,\,\,\,\,\,\,\,  \qquad\qquad \left ( Y_{\{\alpha}^{[a}Y_{\beta \}}^{b]} +  \gamma^\mu_{\alpha\beta}   \Phi^{[i_1} D_\mu \Phi^{i_2]}
\right )  \Phi^{m-2} \end{equation}
The fermionic bilinears transform  in the $[0,0,2,0]$ and $[0,1,0,0]$ representation of SO(8), respectively,  which can be contracted with $[m-2,0,0,0]$ to give the expected representations $[m-2,0,2,0]$ and $[m-2,1,0,0]$ given in Table \ref{S7}.  
The states with dimension  $\frac{m+3}{2}$ read schematically 
\begin{equation}
 \left( Y^a_\alpha Y^b_\beta Y^c_\gamma + \cdots \right )\Phi^{m-3} 
\end{equation}
The Lorentz indices can be completely symmetric  or  with mixed symmetry;  we obtain a state $\frac{3}{2}^{(1)}$ with three SO(8) indices transforming in the $[1,0,0,1]$ and a state $\frac{1}{2}^{(2)}$ with three indices $[0,1,1,0]$, respectively. Combining with the bosonic part we easily recover the representations 
$[m-2,0,0,1]$ and $[m-3,1,1,0]$ reported in Table \ref{S7}. The states with dimension  $\frac{m+4}{2}$ have four fermions whose Lorentz indices can be 
contracted in order to give a scalar $0^{+(2)}$, a vector $1^{+(2)}$ and a spin two $2^{+}$. We can write, for example,  the schematic form of  the spin two operator in the $[m-2,0,0,0]$ representation of SO(8),
\begin{equation}
 D_\mu \Phi^i D_\nu \Phi^{i\, } \,   \Phi^{m-2} + Y_{\{\alpha_1}^{[a_1}\cdots Y_{\alpha_4 \}}^{a_4]}\,  \Phi^{m-4} +\cdots 
\end{equation} 
The scalar and vector are obtained with different contraction of the Lorentz indices; it is easy to check that the SO(8) indices then transform as reported in Table \ref{S7}. The rest of the spectrum can be similarly reconstructed.

The KK spectrum on $S^7$  should reproduce the full spectrum of short operators with finite dimensions in the large N limit of the field theory dual to AdS$_4\times S^7$. It is however difficult to write explicit expressions for these multiplets in terms of local operators. The ABJM theory with $k=1$
is an explicitly scale invariant  local Lagrangian dual to  AdS$_4\times S^7$ and its spectrum of single trace chiral operators should match the KK spectrum on $S^7$.  However, the explicit correspondence is somehow obscured by the fact that the ${\cal N}=8$ supersymmetry of ABJM
is not manifest and monopole operators are required to match the spectrum. In the particular case N$=2$, the ABJM theory with group SU(2)$\times$ SU(2) has manifest ${\cal N}=8$ supersymmetry and SO(8) global symmetry; in fact it coincides with the BLG theory \cite{BGL}. However, even
in this case we can only give an explicit representation for the multiplets with even $m$. Let us discuss briefly how.  We use the formulation as a Chern-Simons theory with gauge group SU(2)$\times$SU(2). The matter content consist of  bosonic and fermionic fields $\Phi^i$ and $Y^a$   transforming  in the bi-fundamental representation of the gauge group and  in the ${\bf 8}_v$ in the ${\bf 8}_s$  representation of SO(8), respectively. The fields can be written as two-by-two matrices satisfying the reality condition
\begin{equation}
\Phi^i_{A\dot A} =\epsilon_{AB}\epsilon_{\dot A\dot B} (\Phi^{i\, \dagger})^{\dot B B}\, , \qquad\qquad (Y^a)_{A\dot A} =\epsilon_{AB}\epsilon_{\dot A\dot B} (Y^{a\,  \dagger})^{\dot B B}
\end{equation}
The reality conditions ensures that the basic degrees of freedom of an ${\cal N}=8$ multiplet consist in eight real bosons and fermions. Compared with the free theory discussed above, the  supersymmetry transformations are modified as 
\begin{eqnarray}
[ Q^{\dot a}_{ \alpha}, \Phi^i ] &=& i (\tilde\Gamma^{i})^{\dot a a} Y^{a}_{\alpha} \nonumber\\
\{Q^{\dot a}_{\alpha}, Y^{a}_{\beta} \}& =&  (\gamma^\mu)_{\alpha\beta} D_\mu \Phi^i (\Gamma^i)^{a\dot a} +
\epsilon_{\alpha\beta} \Phi^i\Phi^{j\, \dagger}\Phi^k (\Gamma^{ijk})^{a\dot a} 
\end{eqnarray}
The lowest state of the $m$-th KK supermultiplet is given by  ${\rm Tr}\,  \Phi^m $ which is schematic expression for a product of $m$ fields $\Phi^{i_k}$ or $\Phi^{i_k\, \dagger }$  completely symmetrized in their SO(8) indices and with all the traces removed
\begin{equation}\Phi^{\{i_1}\Phi^{i_2\, \dagger}\Phi^{i_3}\Phi^{i_4\, \dagger} \cdots \Phi^{i_m \}\, \dagger } -{\rm traces}\end{equation}
In order to have a gauge invariant expression  $m$ must be even and $\Phi$ and $\Phi^\dagger$ should be alternated.
The other states are obtained, similarly to the free case,  by applying the supersymmetry transformations. Let us quote, for example, the schematic form of the  scalar $ 0^{-(1)}$ and the  vector $1^{-(1)}$with dimension  $\frac{m+2}{2}$ 
 \begin{equation}
  {\rm Tr}  \,  \left ( Y_{[ \alpha}^{\{a}Y_{\beta ]}^{b\}\dagger} +  \Phi^{[i_1}\Phi^{i_2\, \dagger}\Phi^{i_3}\Phi^{i_4] \, \dagger}\right )  \Phi^{m-2} \, , \,\,\,\,\,\,\,\,\,    {\rm Tr}  \,  \left ( Y_{\{\alpha}^{[a}Y_{\beta \}}^{b]\dagger} +  \gamma^\mu_{\alpha\beta}   \Phi^{[i_1} D_\mu \Phi^{i_2]\, \dagger}
\right )  \Phi^{m-2} \end{equation}
 
\subsection{The SW spectrum }

The KK spectrum of $S^7/\mathbb{Z}_k$ is also known \cite{Nilsson:1984bj}. $\mathbb{Z}_k$ acts by reducing the length of a circle in $S^7$ and,  for large $k$, the orbifold quotient is equivalent to   a type IIA reduction of the theory along this circle. As a result, for large $k$, the KK spectrum of M theory on $S^7/\mathbb{Z}_k$ is the same as the KK spectrum of type IIA on $\mathbb{CP}^3$.  To obtain the type IIA spectrum we just need to decompose the SO(8)  representations under the residual SU(4)$\times$U(1) group, and project out all 
the states that are not invariant under the U(1) action. This decomposition was exhaustively studied in \cite{Nilsson:1984bj}:
the levels with $m$ odd are completely  projected out, while the levels labelled by even values of $m$ organize themselves into $\mathcal{N}=6$ 
susy multiplets. The resulting spectrum exactly reproduces the set of chiral multiplets of the ABJM theory \cite{Aharony:2008ug} \footnote{This is strictly correct in the $k\rightarrow\infty$ limit. For finite $k$, some Fourier  modes along the M theory circle survive the projection. These states are D$0$
branes in type IIA and are realized in the ABJM model with monopole operators. We will not discuss explicitly such states in the paper.}.

Consider now the skew-whiffed theory. The skew-whiffing transformation corresponds  to a change of chirality for the 
eight dimensional fermion: ${\bf 8}_s \rightarrow {\bf 8}_c$. This transformation can be easily implemented 
on the KK spectrum of $S^7$ given in Table \ref{S7}. We maintain the dimension $\Delta$ invariant while we exchange 
the spinorial indices in the SO(8) representations: $\[a,b,c,d\] \rightarrow \[a,b,d,c\]$. 
After this transformation, we decompose again  the SO(8)  representations under the residual SU(4)$\times$U(1) group, and project out all 
the states that are not invariant under the U(1) action.  For odd $m$ we obtain ``multiplets'' with only fermionic 
degrees of freedom: all the bosons are projected out; for even $m$ we obtain ``multiplets'' 
with only bosonic degrees of freedom: all the fermions are projected out. The result is reported in  Tables \ref{nBos}
and  \ref{nFer}.

\begin{table}[h!]\begin{center}
\begin{tabular}{|c|p{13cm}|c|}
\hline
Spin &  $\hbox{SU(4)}$ & $\Delta$  \\
\hline
$2^+$ & $[n-1,0,n-1]$ & $n+2$ \\
\hline
$1^{-(1)}$ & $[n-1,0,n-1]+[n,0,n]+[n,1,n-2]+[n-2,1,n]$ & $n+1$ \\
\hline
$1^{+}$ & $[n-1,0,n-1]+[n-1,0,n-1]+[n+1,0,n-3]+[n-3,0,n+1]+[n,1,n-2]+[n-2,1,n]+[n-1,1,n-3]+[n-3,1,n-1]+[n-2,2,n-2]$ & $n+2$ \\
\hline
$1^{-(2)}$ & $[n-1,0,n-1]+[n-2,0,n-2]+[n-1,1,n-3]+[n-3,1,n-1]$ & $n+3$ \\
\hline
$0^{+(1)}$ & $[n,0,n]$ & $n$ \\
\hline
$0^{-(1)}$ & $[n-1,0,n-1]+[n+1,0,n-3]+[n-3,0,n+1]+[n,1,n-2]+[n-2,1,n]+[n-1,2,n-1]$ & $n+1$ \\
\hline
$0^{+(2)}$ & $[n-1,0,n-1]+[n,0,n]+[n-2,0,n-2]+[n,1,n-2]+[n-2,1,n]+[n-1,1,n-3]+[n-3,1,n-1]+[n-2,2,n-2]+[n,2,n-4]+[n-4,2,n]$ & $n+2$ \\
\hline
$0^{-(2)}$ & $[n-1,0,n-1]+[n+1,0,n-3]+[n-3,0,n+1]+[n-1,1,n-3]+[n-3,1,n-1]+[n-3,2,n-3]$ & $n+3$ \\
\hline
$0^{+(3)}$ & $[n-2,0,n-2]$ & $n+4$ \\
\hline
\end{tabular}
\end{center}

\caption{Bosonic KK spectrum of the SW theory at level $m=2n$, with $n\ge 1$.  The states with negative Dynkin labels  should be excluded. 
In the case $n=1$ some of the fields are absent, reflecting the fact that the original $S^7$ multiplet is shorter.} 
\label{nBos}
\end{table}
\begin{table}
\begin{center}
\begin{tabular}{|c|p{13cm}|c|}
\hline
Spin &  $\hbox{SU(4)}$ & $\Delta$  \\
\hline
$\frac{3}{2}^{(1)}$ & $[n+1,0,n-1]+[n-1,0,n+1]+[n-1,1,n-1]$ & $n+2$ \\
\hline
$\frac{3}{2}^{(2)}$ & $[n,0,n-2]+[n-2,0,n]+[n-1,1,n-1]$ & $n+3$ \\
\hline
$\frac{1}{2}^{(1)}$ & $[n+1,0,n-1]+[n-1,0,n+1]+[n,1,n]$ & $n+1$ \\
\hline
$\frac{1}{2}^{(2)}$ & $[n+1,0,n-1]+[n-1,0,n+1]+[n,0,n-2]+[n-2,0,n]+[n,1,n]+[n-1,1,n-1]+[n-1,1,n-1]+
[n+1,1,n-3]+[n-3,1,n+1]+[n,2,n-2]+[n-2,2,n]$ & $n+2$ \\
\hline
$\frac{1}{2}^{(3)}$ & $[n,0,n-2]+[n-2,0,n]+[n+1,0,n-1]+[n-1,0,n+1]+[n-1,1,n-1]+[n-1,1,n-1]+[n-2,1,n-2]+
[n+1,1,n-3]+[n-3,1,n+1]+[n-1,2,n-3]+[n-3,2,n-1]$ & $n+3$ \\
\hline
$\frac{1}{2}^{(4)}$ & $[n,0,n-2]+[n-2,0,n]+[n-2,1,n-2]$ & $n+4$ \\
\hline
\end{tabular}
\end{center}
\caption{ Fermionic KK spectrum  of the SW theory at level $m=2n+1$, with $n\ge1$.  The states with negative Dynkin labels  should be excluded. 
In the case $n=1$ some of the fields are absent, reflecting the fact that the original $S^7$ multiplet is shorter.}
\label{nFer}

\end{table}

Let us consider now the non-supersymmetric field theory discussed in the previous Section.  It obviously difficult to predict the spectrum of operators with finite dimensions in a non supersymmetric theory. However, we should at least be able to construct a gauge invariant operator with the right quantum numbers for each of  KK state. We will see now
that the proposed field content is compatible with the distinctive features of the KK spectrum. 
 
The dimensions $\Delta$ of the KK states are all integers. This is compatible with our assumption 
that the elementary fields have classical dimensions (1/2 for the bosons and 1 for the fermions) since, with this assumption, all the gauge invariant operators have integer dimensions. Let us see this in details. Consider a gauge invariant product of $m$ elementary fields.
Consider first the case with  $m$ even. The bosonic operators have integer dimensions: the dual field theory operators are 
obtained with an even number of bosonic fields and an even number of fermionic fields; 
gauge invariant operators of this type are obtained using $X^A$ and 
$\Psi^I$, or $\xi$ fields.   The fermionic operators would have half-integer dimensions: in fact  they would contain an odd number of bosonic fields 
and odd number of fermionic fields. However, in the  proposed field theory it is not possible to build 
up a gauge invariant operator with such field content. We conclude that for even $m$  there are only bosons in the KK  spectrum. 
Consider now the case with $m$ odd. The fermionic operators have integer dimensions: the dual field theory operators 
are obtained with an even number of bosonic fields and an odd number of fermionic fields; 
gauge invariant operators of this type are obtained using 
$X^A$ and $\Psi^I$, or $\xi$ fields.   The bosonic operators instead would have half-integer dimensions, in fact they would contain an odd number of bosonic  fields and an even number of fermionic fields. Such bosonic operators are still forbidden by gauge invariance.
We conclude that for odd $m$  there are only fermions in the spectrum.

Let us now check that we can construct at least one field theory operator with the right dimension, Lorentz and SU(4) representation for each  KK mode in supergravity. 
The gauge invariant  field theory operators are obtained by contracting the following operators
$$\{ X^A, X^{\dagger}_A, \xi_{\alpha}, \xi^{\dagger}_{\alpha}, \Psi^I_{\alpha}, D_{\mu}\}$$
Let us start looking to the generic form of the operators of low dimension for a given $n$. 
The first scalar KK modes is $0^{+(1)}$ and the corresponding field theory operator has the schematic  form
\begin{equation}\label{scalarBPS}
\Tr \( X^{\{ A_1}X_{\{B_1}^{\dagger} ....... X^{A_{n}\}}X_{B_{n}\}}^{\dagger} \) 
\end{equation}
where suitable trace subtractions are understood.  The first KK spinor mode is $\frac{1}{2}^{(1)}$ and the dual operator is:
 $$\Tr \( X^{\{ A_1}X_{\{B_1}^{\dagger} ....... X^{A_{n}\}}X_{B_{n}\}}^{\dagger} \Psi^I_{\alpha} \) $$
The first KK vector mode is $1^{(-1)}$  and the dual operator is:
$$\Tr \( X^{\{ A_1}X_{\{B_1}^{\dagger} ....... X^{A_{n-1}\}}X_{B_{n-1}\}}^{\dagger} \Psi^{ [ I}_{ \{ \alpha}  \Psi^{J ]}_{\beta \}}\) +  \Tr \( X^{\{ A_1}X_{\{B_1}^{\dagger} ....... X^{A_{n-1}\}}X_{B_{n-1}\}}^{\dagger} \xi_{ \{ \alpha}  \xi^{\dagger}_{\beta \}}\) + $$
$$\Tr \( X^{\{ A_1}X_{\{B_1}^{\dagger} ....... X^{A_{n-1}\}}X_{B_{n-1}\}}^{\dagger}  X^{A_{n}} D_{\mu}X_{B_{n}}^{\dagger}  \)$$
The first pseudo-scalar in the KK spectrum is  $0^{(-1)}$ and the dual field operator is schematically: 
$$\Tr \( X^{\{ A_1}X_{\{B_1}^{\dagger} ....... X^{A_{n}\}}X_{B_{n}\}}^{\dagger} \Psi^{ \{ I}_{ [ \alpha}  \Psi^{J \}}_{\beta ]}\) + \Tr \( X^{\{ A_1}X_{\{B_1}^{\dagger} ....... X^{A_{n-1}\}}X_{B_{n-1}\}}^{\dagger} \xi_{ [ \alpha}  \xi^{\dagger}_{\beta ]}\) + $$
$$\Tr \( X^{\{ A_1}X_{\{B_1}^{\dagger} ....... X^{A_{n}\}}X_{B_{n}\}}^{\dagger}  X^{[ A} X^{\dagger}_{[B }X^{C]} X_{D ] }^{\dagger}  \)$$
Another interesting operator of lower dimension in the KK towers is $ \frac{3}{2}^{(1)}$ and its dual field operator has the form:
$$ \Tr \( X^{\{ A_1}X_{\{B_1}^{\dagger} ....... X^{A_{n-1}\}}X_{B_{n-1}\}}^{\dagger} \Psi^{ [ I}_{ \{ \alpha}  \Psi^{J }_{\beta }\Psi^{K ] }_{\gamma \} }\) + \Tr \( X^{\{ A_1}X_{\{B_1}^{\dagger} ....... X^{A_{n-1}\}}X_{B_{n-1}\}}^{\dagger} \Psi^{ I}_{ \{ \alpha}  \xi_{\beta } \xi_{\gamma \} }^{\dagger} \) +$$
$$ \Tr \( X^{\{ A_1}X_{\{B_1}^{\dagger} ....... X^{A_{n}\}}X_{B_{n}\}}^{\dagger}  D_{\mu} \Psi_{\alpha}^I \)$$


It is easy to verify that the above operators can reproduce the SU(4) representations reported in Tables \ref{nBos} and \ref{nFer}. 
The first bosonic level has $m=2$ and the corresponding operators have the schematic form:
\begin{eqnarray}
0^{+(1)} &:&  \Tr\( X^A X_B^{\dagger} \) \nonumber \\
 0^{-(1)} &:& \Tr \( \Psi^{ \{ I}_{ [ \alpha}  \Psi^{J \}}_{\beta ]}\) 
 + \Tr \(\xi^{\dagger}_{[\alpha} \xi_{\beta]} \)+  \Tr\( X^{[A} X_{[B}^{\dagger} X^{C]} X^{\dagger}_{ D]} \) \nonumber \\  
1^{-(1)} &:&
\Tr\(X^A  D_{\mu} X_B^{\dagger}\) + \Tr \(\xi^{\dagger}_{\{\alpha} \xi_{\beta\}} \) + \Tr \( \Psi^{ [ I}_{ \{ \alpha}  \Psi^{J ]}_{\beta \} }\) \nonumber \\
2^{+} &:& 
\Tr\( D_{\mu}X^A D_{\nu}X_A^{\dagger}\) + ... \nonumber \\  
\end{eqnarray}
where suitable trace subtractions are understood. In the first bosonic level there are two important operators:  the part of the $1^{-(1)}$ operator  in the $\[1,0,1\]$ representation is the field theory SU(4) global current symmetry $J^{[IJ]}_{\mu}$ dual to the gauge field $A^{[IJ]}_{\mu}$ in the adjoint representation of SU(4) propagating in AdS$_4$, and  the singlet $2^{+}$ operator is the stress-energy tensor  dual to  the graviton $g_{\mu \nu}$ in AdS$_4$. Of course, since the proposed skew-whiffed theory is not supersymmetric, the supercurrent operator dual to the would be $\frac{3}{2}^{(1)}$ gravitino  is not present in the operator spectrum. 

The first fermionic level has $m=3$ and the corresponding operators are \footnote{The operators corresponding $\frac{1}{2}^{(2)}$ have a mixed symmetry in the spin and SU(4) indices which is indicated in a somehow imprecise way in the following formula.}: 

\begin{eqnarray}
\frac{1}{2}^{(1)} &:&
 \Tr\( X^A X_B^{\dagger} \Psi^I_{\alpha} \) \nonumber \\
 \frac{1}{2}^{(2)} &:& 
\Tr\( \Psi^{ [ I}_{\{ \alpha} \Psi^{ \{J ] }_{ [ \beta\} } \Psi^{K\} }_{\gamma ] } \)   +     \Tr\( \Psi^{I}_{\{ \alpha} \xi_{ [ \beta \} } \xi^{\dagger}_{\gamma ] } \)  + \Tr\( X^{ [ A} X_{ [B}^{\dagger} X^{C ]} X^{\dagger}_{D]} \Psi^I_{\alpha}\) \nonumber\\ 
\frac{3}{2}^{(1)} &:&
  \Tr\( \Psi^{ [ I}_{\{ \alpha} \Psi^{J}_{\beta} \Psi^{K ]}_{\gamma \} } \) +   \Tr\( \Psi^{I}_{\{ \alpha} \xi_{\beta} \xi^{\dagger}_{\gamma \} } \) +  \Tr\( X^A X_B^{\dagger} D_{\mu} \Psi^I\) \nonumber \\
\frac{3}{2}^{(2)} &:&
 \Tr\( \xi_{ [ \alpha} \xi^{\dagger}_{\beta] }  D_{\mu} \Psi_{\gamma}^I \) + ...
\end{eqnarray}

The operators dual to a specific KK modes are, in general, linear combinations of all the gauge invariant operators  
 with the right dimension, Lorentz and SU(4) representation that we can construct. The expectation is that exactly one of these operators will remain with finite dimensions in the large N and strongly coupled regime, while the others will acquire infinite dimension, as standard in the AdS/CFT correspondence.

\section{Conclusions}\label{conc}

In this paper we discussed general properties of  the field theory living on a stack of N anti-M2 branes at the tip of an eight dimensional real cone with at least one Killing spinor.  In particular we focus on the skew-whiffed AdS$_4\times S^7/\mathbb{Z}_k$ supergravity solution in M theory (and its reduction to type IIA), which is not supersymmetric but  stable.  The AdS/CFT correspondence predicts the existence of a dual non-supersymmetric three-dimensional  theory,  which should be a unitary conformal field theory at least at large N and strong coupling.  Based on an analysis of the KK spectrum of the supergravity theory and on the geometrical action of the skew-whiffing transformation, we speculate on the field content of the dual theory, a  skew-whiffed version of the ABJM theory. Due to the fact that the theory is non-supersymmetric, direct checks of any proposal  are non trivial. We identified  a field content which is at least compatible with the KK spectrum found in the dual supergravity solution. 
It is interesting to observe that there is some ambiguity in writing a candidate theory. In particular  the SW map and the compatibility with the KK spectrum do not fix unambiguously the representation of the fields under the gauge group; 
for example,  the complex spinor $\xi$ could be either in the bifundamental representation, as we considered in this paper, or in the adjoint representation of one of the gauge group. It would be interesting to see if these various theories are somehow duals.  Another observation is that the proposed theory is not really a quiver gauge theory, because the fermions transform in the bifundamental of just the SU(N) part of the full U(N) gauge group. This seems to implies that the theory is a good candidate only in the large N limit, that is exactly, the regime for which the dual supergravity solution is demonstrated to be stable. Further investigations are needed. 

We mainly concentrated on  the $S^7/\mathbb{Z}_k$ supergravity solution. However, as explained at the very beginning of the paper, there is a SW supergravity solution for  any Freud-Rubin solution with at least one Killing spinor and we should be able to find a dual non  supersymmetric Chern-Simons matter theory. There is an infinite set of AdS$_4$/CFT$_3$ pairs in the literature and it would be interesting to  have a systematic procedure to obtain the SW field theory once the supersymmetric field theory is known. One way to proceed would be to start with  orbifolds of the SW ABJM theory.  By giving vevs to scalar fields we can  flow to other  non-supersymmetric Chern-Simons matter field theories.  A more efficient way to proceed would be to find the field theory operation dual to the change of orientation of H$_7$. We leave this topic for future research.

\section*{Acknowledgements}

We would like to thank J. Gauntlett, N. Halmagyi, A. Hanany, S. Hartnoll,  N. Lambert, A. Sen, Y. Tachikawa, J. Troost, K. Zarembo for nice and helpful discussions. 
The work of D.F. is partially supported by IISN - Belgium (convention 
4.4514.08), by the Belgian Federal Science Policy Office through the 
Interuniversity Attraction Pole P6/11 and by the ``Communaut\'e 
Francaise de Belgique" through the ARC program. A.~Z.~ is supported in part by INFN.

\end{document}